%% file: dilepton_prl.tex
\documentstyle[preprint,eqsecnum,epsf,aps,floats,tighten]{revtex}
\def\met{\mbox{${\hbox{$E$\kern-0.6em\lower-.1ex\hbox{/}}}_T$}}
\def\squark{\mbox{$\widetilde{q}$}}
\def\gluino{\mbox{$\widetilde{g}$}}
\def\chargino{\mbox{$\widetilde{\chi}^\pm_{1-2}$}}
\def\neutralino{\mbox{$\widetilde{\chi}^0_{1-4}$}}

\begin{document}


\title{ A Search for Dilepton Signatures from Minimal Low-Energy
Supergravity in $p\overline{p}$ Collisions at $\sqrt{s}=1.8$~TeV }

%
\author{                                                                      
B.~Abbott,$^{46}$                                                             
M.~Abolins,$^{43}$                                                            
V.~Abramov,$^{19}$                                                            
B.S.~Acharya,$^{13}$                                                          
D.L.~Adams,$^{53}$                                                            
M.~Adams,$^{30}$                                                              
S.~Ahn,$^{29}$                                                                
V.~Akimov,$^{17}$                                                             
G.A.~Alves,$^{2}$                                                             
N.~Amos,$^{42}$
% ...
\\                                                                            
\vskip 0.30cm                                                                 
\centerline{(D\O\ Collaboration)}                                             
\vskip 0.30cm                                                                 
}                                                                             
\address{                                                                     
\centerline{$^{1}$Universidad de Buenos Aires, Buenos Aires, Argentina}       
\centerline{$^{2}$LAFEX, Centro Brasileiro de Pesquisas F{\'\i}sicas,         
                  Rio de Janeiro, Brazil}                                     
\centerline{$^{3}$Universidade do Estado do Rio de Janeiro,                   
                  Rio de Janeiro, Brazil}                                     
\centerline{$^{4}$Institute of High Energy Physics, Beijing,                  
                  People's Republic of China}                                 
\centerline{$^{5}$Universidad de los Andes, Bogot\'{a}, Colombia}             
\centerline{$^{6}$Universidad San Francisco de Quito, Quito, Ecuador}         
\centerline{$^{7}$Institut des Sciences Nucl\'eaires, IN2P3-CNRS,             
                  Universite de Grenoble 1, Grenoble, France}                 
\centerline{$^{8}$Centre de Physique des Particules de Marseille,             
                  IN2P3-CNRS, Marseille, France}                              
\centerline{$^{9}$LPNHE, Universit\'es Paris VI and VII, IN2P3-CNRS,          
                  Paris, France}                                              
\centerline{$^{10}$DAPNIA/Service de Physique des Particules, CEA, Saclay,    
                  France}                                                     
\centerline{$^{11}$Panjab University, Chandigarh, India}                      
\centerline{$^{12}$Delhi University, Delhi, India}                            
\centerline{$^{13}$Tata Institute of Fundamental Research, Mumbai, India}     
\centerline{$^{14}$Seoul National University, Seoul, Korea}                   
\centerline{$^{15}$CINVESTAV, Mexico City, Mexico}                            
\centerline{$^{16}$Institute of Nuclear Physics, Krak\'ow, Poland}            
\centerline{$^{17}$Institute for Theoretical and Experimental Physics,        
                   Moscow, Russia}                                            
\centerline{$^{18}$Moscow State University, Moscow, Russia}                   
\centerline{$^{19}$Institute for High Energy Physics, Protvino, Russia}       
\centerline{$^{20}$Lancaster University, Lancaster, United Kingdom}           
\centerline{$^{21}$University of Arizona, Tucson, Arizona 85721}              
\centerline{$^{22}$Lawrence Berkeley National Laboratory and University of    
                   California, Berkeley, California 94720}                    
\centerline{$^{23}$University of California, Davis, California 95616}         
\centerline{$^{24}$California State University, Fresno, California 93740}     
\centerline{$^{25}$University of California, Irvine, California 92697}        
\centerline{$^{26}$University of California, Riverside, California 92521}     
\centerline{$^{27}$Florida State University, Tallahassee, Florida 32306}      
\centerline{$^{28}$University of Hawaii, Honolulu, Hawaii 96822}              
\centerline{$^{29}$Fermi National Accelerator Laboratory, Batavia,            
                   Illinois 60510}                                            
\centerline{$^{30}$University of Illinois at Chicago, Chicago,                
                   Illinois 60607}                                            
\centerline{$^{31}$Northern Illinois University, DeKalb, Illinois 60115}      
\centerline{$^{32}$Northwestern University, Evanston, Illinois 60208}         
\centerline{$^{33}$Indiana University, Bloomington, Indiana 47405}            
\centerline{$^{34}$University of Notre Dame, Notre Dame, Indiana 46556}       
\centerline{$^{35}$Iowa State University, Ames, Iowa 50011}                   
\centerline{$^{36}$University of Kansas, Lawrence, Kansas 66045}              
\centerline{$^{37}$Kansas State University, Manhattan, Kansas 66506}          
\centerline{$^{38}$Louisiana Tech University, Ruston, Louisiana 71272}        
\centerline{$^{39}$University of Maryland, College Park, Maryland 20742}      
\centerline{$^{40}$Boston University, Boston, Massachusetts 02215}            
\centerline{$^{41}$Northeastern University, Boston, Massachusetts 02115}      
\centerline{$^{42}$University of Michigan, Ann Arbor, Michigan 48109}         
\centerline{$^{43}$Michigan State University, East Lansing, Michigan 48824}   
\centerline{$^{44}$University of Nebraska, Lincoln, Nebraska 68588}           
\centerline{$^{45}$Columbia University, New York, New York 10027}             
\centerline{$^{46}$New York University, New York, New York 10003}             
\centerline{$^{47}$University of Rochester, Rochester, New York 14627}        
\centerline{$^{48}$State University of New York, Stony Brook,                 
                   New York 11794}                                            
\centerline{$^{49}$Brookhaven National Laboratory, Upton, New York 11973}     
\centerline{$^{50}$Langston University, Langston, Oklahoma 73050}             
\centerline{$^{51}$University of Oklahoma, Norman, Oklahoma 73019}            
\centerline{$^{52}$Brown University, Providence, Rhode Island 02912}          
\centerline{$^{53}$University of Texas, Arlington, Texas 76019}               
\centerline{$^{54}$Texas A\&M University, College Station, Texas 77843}       
\centerline{$^{55}$Rice University, Houston, Texas 77005}                     
\centerline{$^{56}$University of Washington, Seattle, Washington 98195}       
}                                                                             

\date{\today}
\maketitle
\vskip 0.5cm

\newpage
\begin{abstract}
\noindent
We report on a search for supersymmetry  using the D\O\ detector. The
1994-1996 data sample of $\sqrt{s}=1.8$~TeV $p\overline{p}$  collisions
 was analyzed for events containing
two leptons ($e$ or $\mu$), two or more jets, and missing transverse
energy. Assuming the minimal supergravity model, with 
$A_0=0$ and $\mu<0$, various thresholds
 were employed to
optimize the search. No events were found beyond expectation
from background. We set a lower limit
at the 95\% 
C.L. of 255  GeV/$c^2$ for equal mass squarks and gluinos for $\tan{\beta}=2$, 
and present exclusion contours in the $(m_0,m_{1/2})$
plane for $\tan{\beta}=2$--6.

\end{abstract}


      Supersymmetric extensions of the standard model \mbox{(SM)} have been
      the subject
      of intense theoretical and experimental investigation in  recent years.
      The simplest, the minimal supersymmetric standard model (MSSM),
      incorporates supersymmetry (SUSY) \cite{theory}, a fundamental 
      space-time symmetry relating 
      fermions to bosons. SUSY requires the existence 
      of a partner (a sparticle) for
      every SM particle, and at least one additional Higgs doublet.  The added
      assumption of conservation of $R$-parity,
      a multiplicative quantum number ($+1$ for SM particles and $-1$ for
       their SUSY counterparts),
      implies the pair production of sparticles in high energy
      collisions. The sparticles can decay directly, or via
       lighter 
      sparticles, into final states
      containing SM particles and stable  lightest supersymmetric
      particles (LSPs).
      LSPs are weakly interacting objects \cite{lsp} that escape detection,
      and produce a large apparent imbalance in transverse energy ($\met$) 
      in the event. This is a characteristic
      signature for SUSY processes.
    
      In this Letter we describe a search for production of squarks
  (\squark), gluinos (\gluino), charginos (\chargino), and/or  neutralinos
  (\neutralino). Cascade decays of 
      these sparticles can have significant leptonic branching fractions.
      For example, \gluino\ cascades can terminate with the decay 
    $\widetilde{\chi}^0_2\rightarrow l\overline{l} \widetilde{\chi}^0_1$
    25\% 
    of the time \cite{thesis}. We 
    consider final states containing two isolated 
      leptons ($e$ or $\mu$), two or more jets (or three or more jets), 
    and \met \cite{thesis}, 
      thus complementing searches that consider only
      jets and \met\ \cite{ref:met}.

    Because of the large number of free parameters in the generic MSSM, we have
    chosen to compare our data with a class of minimal low-energy supergravity
    (mSUGRA) models \cite{sugra} that are more tightly constrained. These are 
      parameterized in terms of only five free parameters:
       a common SUSY-breaking mass ($m_{0}$) for all scalars, a common mass
      for all gauginos ($m_{1/2}$), a common value for all trilinear couplings
      ($A_0$), the ratio of the vacuum expectation values of the two Higgs 
     fields (tan$\beta$),
      and the sign of $\mu$, where $\mu$ is the Higgsino mass parameter.
     We assume $A_0=0$ and $\mu<0$ in this analysis.

      The D\O\ detector \cite{ref:d0} consists of a  
     liquid-argon calorimeter surrounding central tracking chambers, all
      enclosed within an
      iron toroidal muon spectrometer. 
	Structurally, the calorimeter is segmented into a central
       calorimeter (CC)
	and two end calorimeters (EC). 
    Within the central tracking
    chambers, a transition radiation detector (TRD) aids in
    electron identification in the CC.

The data were collected during the 1994-1996 Tevatron collider run.
 We triggered
on an electron, one jet, and \met\, for the $ee$ and $e\mu$ signatures,
 and on a
muon and a jet for the $\mu \mu$ signatures. The integrated luminosity was
$108\pm6$~pb$^{-1}$ for $ee$ and $e\mu$ signatures,
 and $103\pm5$~pb$^{-1}$ for $\mu\mu$
signatures.     The original data sample of several million events was
reduced by requiring that events have two
    leptons satisfying loose identification criteria, two jets with 
    $E_T>15$~GeV, and \met$>14$~GeV. 
This sample of 24,233 predominantly multijet events 
was used in the subsequent analysis.

    Jets were reconstructed from the energy deposition 
    in the calorimeter in cones of radius 
    ${\cal R}=\sqrt{(\Delta\eta)^2+(\Delta\phi)^2}=0.5$, where $\phi$ is the
 azimuthal
    angle with respect to the beam axis, and $\eta$ is the pseudorapidity. 
Additional details concerning 
     reconstruction and energy calibration can be found in Refs.
    \cite{ref:d0,ref:jes,ref:topprd}. Jets were required to be in the region
	$|\eta|<2.5$.



 We selected electrons in the CC ($|\eta| < 1.1$) and in the EC
      ($1.5 < |\eta| < 2.5$) using, respectively, a 5-variable and a 4-variable
      likelihood function based  on
    the fraction of energy deposited
    in the electromagnetic (EM) portion of the calorimeter,
    a shower-shape variable, 
    track ionization ($dE/dx$) in the central detector, 
    the quality of the match between the
    reconstructed track and the center of gravity of the 
    calorimeter cluster
    ($\sigma_{\mathrm TRK}$), and a
    variable based on the energy deposited in the TRD
    (not used for the EC). The
    identification efficiency for electrons was determined using a sample of
    $Z\rightarrow ee$ events, and depends on jet multiplicity (high
    track-multiplicity degrades the resolution of
    $\sigma_{\mathrm TRK}$). We defined an electron isolation variable 
${\cal I}=(E_{\mathrm tot}^{\mathrm 0.4} - E_{\mathrm EM}^{\mathrm 0.2})/ (E_{\mathrm EM}^{\mathrm 0.2})$,
    where $ E_{\mathrm EM}^{\mathrm 0.2}$ is the EM energy in a cone of
 ${\cal R}=0.2$,
    and $E_{\mathrm tot}^{\mathrm 0.4}$ is the total calorimeter energy in a
 cone of ${\cal R}=0.4$. 
    We required ${\cal I}<0.3$ in this analysis. The identification
 efficiencies for
    isolated electrons were typically 78--84\% for CC electrons, and 
    63--69\% for EC electrons \cite{thesis}. 


    Muon identification is detailed in Ref. \cite{ref:topprd}. 
    Muons were required to have $|\eta|<1.7$ and to lie outside
    of all reconstructed jets defined by ${\cal R}=0.5$ cones.
 To remove poorly measured
    muons, the direction of the vector \met\ was required to be more
    than 10 degrees in $\phi$ away 
    from any muon track; this reduced the acceptance by about 10\% per muon.


    Our data sample was further refined by requiring two good jets 
    with $E_T>20$ GeV, \met$>20$ GeV, a fiducial cut on the 
    event vertex \cite{thesis}, and offline lepton 
    selections of:
    $E_T(e_1)>17$ GeV and $E_T(e_2)>15$ GeV, or $E_T(e)>17$ GeV and 
    $E_T(\mu)>4$ GeV, 
    or $E_T(\mu_1)>20$ GeV and 
    $E_T(\mu_2)>10$ GeV. 
This left 10 $ee$, 6 $e\mu$, and 
3~$\mu\mu$ events.

    Background came from four sources: $t\overline{t}$, $Z$ and
 $W$ boson, and QCD jet
    production. 
    The $t\overline{t}$ and $Z$ boson backgrounds were calculated using
 published cross
    sections \cite{ref:topxsect,ref:wxsprd} and a
    fast detector-simulation package (described below),
      while QCD multijet and $W$+jets backgrounds were estimated 
      from data.
      For the $ee$ and $e\mu$ signatures, we selected events with nearly
      the same topology, 
      except that one isolated electron was missing and an extra jet
      was required in its place.
    The background was then estimated  using the measured
    probability of one of the jets being misidentified as an isolated
    electron \cite{thesis}.
    For $\mu\mu$ signatures, the background sample was defined by one
 isolated and one
    non-isolated muon (within a jet), and two or three other jets.  The
    measured probability for a non-isolated muon to appear as an isolated
    muon was used to estimate the background from this
 source \cite{ref:topprd}.
      The QCD and $W$+jets backgrounds were combined because
      they are topologically similar: for $W$ boson events, the identified
      lepton is real, and for QCD the identified lepton is due to a jet
      fluctuation. For the accepted $ee$ and $\mu\mu$ events, about 50\% of the
      background results  from $Z$ boson production, 30\% from QCD/$W$+jets, 
	and 20\% from
      $t\overline{t}$ production. For the accepted $e\mu$ events, 
    the breakdown was 10\%, 60\%, and 30\%, respectively. 

    The uncertainties in the QCD/$W$+jets backgrounds stemmed from 
    energy scale (12\%), the probability of lepton
     misidentification (15\%), and
    statistics (2--100\%). The uncertainties in the other
     backgrounds were due to
    trigger and identification efficiencies (11--15\%), 
    cross section (8--30\%),
    energy scale (2\%), and Monte Carlo statistics (2--50\%). The large
    statistical uncertainties dominate only when backgrounds 
    are negligible ($<0.1$ events).

\begin{figure}
      \begin{center}
      \mbox{}
      \epsfxsize = 8.0cm \epsfbox{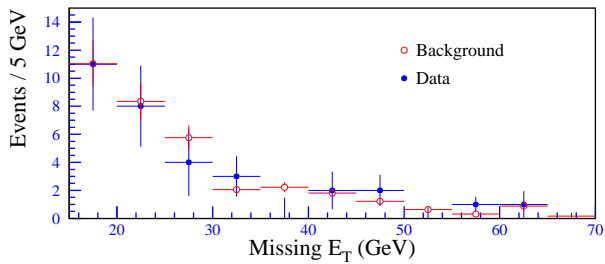}
      \end{center}
    \vskip -0.5cm
      \caption{Comparison of the \met\ distributions for data and
       background for
      $ee+1$-jet events (see text).}
    \label{fig:met}
      \end{figure}

    To check for systematic uncertainties in misidentification of electrons,
    we enlarged our $ee$ event sample by 32 events by selecting
   interactions that
    contained two good electrons and at least one jet. The \met\ for these 42 
    events is compared in Fig.~\ref{fig:met} with the analogous background 
    estimate from  QCD/$W$+jets.
    The two distributions in Fig.~\ref{fig:met} were normalized to each
    other in
    the 15--20 GeV interval, where background dominates, and are seen to be
    consistent over the entire range of \met, thereby supporting an assertion
    that the selected $ee$ events are consistent with mismeasurement (or  
    fluctuation) of energy in the calorimeter.

      The usual way to search for a signal is to generate signal and background
  events and then to 
      optimize a single set of requirements that yields the best
 discrimination. 
      A problem with this method is that the optimum thresholds vary
      as a function of the mSUGRA input parameters. In essence, one must select
      different 
      requirements at every point in model space, which is exceptionally
      demanding in computing resources. 

    We performed  an approximate optimization of selection criteria on a grid
    of thresholds, as exemplified in Table~\ref{tb:results}.  
    For $ee$ signatures, we considered sets of
   requirements both with and without an exclusion of $ee$ invariant mass
   ($M_{ee}$) around the $Z$ boson mass.
      For $\mu \mu$ signatures, 
      a cut of \met$>40$~GeV provided the best reduction in the $Z$ boson
 background. Each unique
      combination of thresholds is called a {\em channel}.
      In all, we defined 16 $ee$, $24$ $e\mu$, and 12 $\mu \mu$ channels, 
	for a total of 52.

      To handle the large number of channels, a specialized Monte Carlo was
      written \cite{thesis} that incorporated 
    {\sc spythia} \cite{ref:spythia} as the event 
      generator. This Monte Carlo used a fast simulation of  the detector, 
the trigger, and particle
      identification, 
      using efficiencies and resolutions from  data, and calculated the
      probability of observing events in each of the 52 
      channels. The primary outputs were the efficiencies
$\epsilon_i = B \cdot \varepsilon_{\mathrm trig} \cdot \varepsilon_{\mathrm id} \cdot a_{\mathrm det}$
(products of the branching fraction, trigger efficiency, identification
 efficiency, and detector acceptance, respectively)
for each channel $i$, and the theoretical production cross
section. The fast Monte Carlo reproduced efficiencies obtained in 
a more detailed simulation to 1--2\% accuracy.

Because looser requirements produced event samples that were
 supersets of tighter
requirements, the channels within a 
given signature are correlated. To
avoid bias, we chose a ``best'' channel for each signature (repeated for each
mSUGRA model analyzed) based on the background estimate and
expected signal. Specifically, for each model $k$, 
where $k$ denotes a specific choice of 
$m_0$, $m_{1/2}$ and $\tan{\beta}$, 
we defined an
 expected significance for channel $i$:
$ \overline{S}_i^k = \sum_{N=0}^\infty P(s_i^k+b_i|N) \cdot S(b_i|N)$,
where $P$ is the Poisson probability that signal, $s_i^k$, and
 background, $b_i$,
produce $N$ observed events, and $S$ is the Gaussian significance, 
{\em i.e.} the number of standard deviations that
background must fluctuate to produce $N$ events \cite{ref:sig}. 
Clearly, the sensitivity of the search, as reflected in the above sum over 
all $N$ possible outcomes of the experiment, improves when the probabilities 
$P(s_i^k+b_i|N)$ are sizeable, but the likelihoods of $b_i$ fluctuating 
to $N$ are small ({\em i.e.}, $S(b_i|N)$ are large).
The three maximum $\overline{S}_i^k$ values define 
three independent optimized search channels: $ee_{\mathrm best}^k$, 
$e\mu_{\mathrm best}^k$, and $\mu\mu_{\mathrm best}^k$. The single
 best of the two- or three-
channel combinations, ($cmb_{\mathrm best}^k$), is again defined  
by the analogous maximum $\overline{S}^k_{cmb}$, yielding 
four search channels per model. 

For each model $k$,
we calculated the four 95\% confidence level (C.L.) limits
 on the cross section,
$\sigma^x_{{\mathrm lim}}$ with $x=ee$, $e\mu$, $\mu\mu$, or $cmb$,  
using a standard Bayesian prescription, with a flat prior
for the signal cross section \cite{thesis}.
We also calculated a model-independent limit for the product
$\epsilon\cdot\sigma$.
Table~\ref{tb:results} 
summarizes the background predictions and the number of observed
events in representative  channels, and  the (one-sided Poisson)
 probability that
the background fluctuated to produce the observed events. Indicated in 
{\bf bold} font are
the three best channels for the model $m_0=280$ GeV/$c^2$ and
 $m_{1/2}=51$ GeV/$c^2$, 
where we obtained
$\epsilon^{ee}=(0.49\pm 0.05)$\%, 
$\epsilon^{e\mu}=(0.09\pm 0.01)$\%,
$\epsilon^{\mu\mu}=(0.24\pm 0.02)$\%, and 
$\sigma^{ee}_{{\mathrm lim}}=58$~pb, for 
$\sigma_{{\mathrm tot}}=84$~pb.

\begin{figure}
      \begin{center}
      \mbox{}
      \epsfxsize = 9.5cm \epsfbox{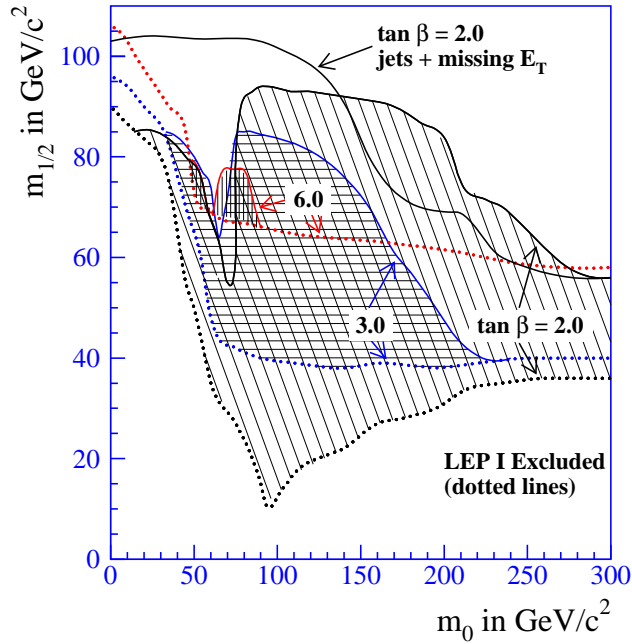}
      \end{center}
      \caption{The hatched regions are excluded
	by the dilepton search at the 95\%
      C.L. for $\tan{\beta}=2$ (diagonal), 3 (horizontal), and 6
      (vertical), with $A_0=0$ and $\mu<0$. The regions below the
	dotted lines are excluded by LEP~I. The result
         from the D\O\ jets and 
	 ${\hbox{$E$\kern-0.6em\lower-.1ex\hbox{/}}}_T$
	search \protect \cite{ref:met} is also shown.}
    \label{fig:d0dilep}
      \end{figure}

      We generated  about 10,000 models, $k$, 
      randomly in the space
      $0<m_0<300$~GeV/$c^2$,  $10<m_{1/2}<110$~GeV/$c^2$, and
      $1.2<\tan{\beta}<10$, to
      obtain a rough exclusion region. Near the boundary of the 
      $m_0$ and $m_{1/2}$ exclusion region, higher statistics samples
    were generated for several values of $\tan{\beta}$.
      Figure~\ref{fig:d0dilep} shows the 95\% C.L. exclusion regions for
    $\tan{\beta}=2$, 3, and 6. Published results from LEP~I \cite{ref:lep1}
     and D\O\ for the jets + \met\ channel \cite{ref:met} are shown  for
    comparison.
    For $\tan{\beta}>6.0$, we do not exclude models not
    previously excluded by LEP~I. (Recent results from LEP~II \cite{ref:lep2}
    provide limits comparable to those presented 
    in this Letter.)



    The contours in Fig.~\ref{fig:d0dilep} have structure that can be 
	understood as follows. 
    First, the dip near 
    $m_0=80$~GeV/$c^2$ for $\tan{\beta}=2.0$ is caused by the
    dominance of the decay 
    $\widetilde{\chi}_2^0\rightarrow \nu \overline{\nu} \widetilde{\chi}_1^0$
    over 
    $\widetilde{\chi}_2^0\rightarrow l^{+} l^{-} \widetilde{\chi}_1^0$
    in this region of phase space. 
    Sensitivity improves for  
    $\tan{\beta}$ closer to 3.0 due to several factors:
    gaugino mass couplings increase, 
    causing the $\widetilde{\chi}_2^0$ to preferentially decay into quarks and
    become a source of 
    jets; gaugino masses decrease,  and decays of 
    squarks into $\widetilde{\chi}_3^0$ and $\widetilde{\chi}_4^0$ become
    allowed; $\widetilde{\chi}_3^0$ and $\widetilde{\chi}_4^0$ 
    dominantly decay into sneutrinos, $\widetilde{\nu}_l$; 
    $\widetilde{\nu}_l\rightarrow \widetilde{\chi}_1^\pm l^\mp$ dominates in 
    this region and becomes a source of leptons. Sensitivity decreases again
    for  $\tan{\beta}$ values around 6.0, where
    decays into light charged leptons are reduced by increased couplings to
    large mass fermions.
    Second, the exclusion for $m_{1/2}$ decreases for large $m_0$, which
    corresponds  to the region where 
    $m_{\tilde{q}}\gg m_{\tilde{g}}$,  and squark
    production does not contribute. In this asymptotic region, we
    exclude gluinos with mass below 
    175~GeV/$c^2$ for $\tan{\beta}=2.0$.
    For squarks and gluinos of equal mass, we exclude masses below 
    255~GeV/$c^2$ for $\tan{\beta}=2.0$.
    We also
    exclude gluinos below 129~GeV/$c^2$ and squarks below 138~GeV/$c^2$,  for
    $m_0<300$~GeV/$c^2$ and $\tan{\beta}<10.0$. 

    In conclusion, we have performed a search for dilepton signatures from
    squark, gluino, and gaugino production. No significant excess of events
    was observed
    and we have presented our results in terms of contours of 
    of exclusion  in mSUGRA parameter space.


\input{table1.tex}

%
We thank C. Kolda, S. Mrenna, G. Anderson, J. Wells, and G. L.
Kane for helpful discussions.
We thank the staffs at Fermilab and at collaborating institutions 
for contributions to this work, and acknowledge support from the 
Department of Energy and National Science Foundation (USA),  
Commissariat  \` a L'Energie Atomique (France), 
Ministry for Science and Technology and Ministry for Atomic 
   Energy (Russia),
CAPES and CNPq (Brazil),
Departments of Atomic Energy and Science and Education (India),
Colciencias (Colombia),
CONACyT (Mexico),
Ministry of Education and KOSEF (Korea),
CONICET and UBACyT (Argentina),
A.P. Sloan Foundation,
and the Humboldt Foundation.

\end{document}

%% file: table1.tex
\begin{table}
\begin{center}
\begin{tabular}{dccccddd}
\multicolumn{8}{c}
{    Signature: $ee+{\mathrm jets}+$ \met }\\ \tableline 

$j_1$ & \multicolumn{2}{c}{$N_{\mathrm jets}$} & \met & 
Background & Data & Prob. (\%) &
$(\epsilon\sigma)_{\mathrm lim}$  \\ 
\tableline
20 &\multicolumn{2}{c}{ 2} & 20 & 
     10.67 $\pm$ 1.24 &  10 & 50.1 & 85\\
20 &\multicolumn{2}{c}{ 3} & 20 & 
     3.08 $\pm$ 0.39 & 2 & 40.3 & 42\\
20 &\multicolumn{2}{c}{ 3} & 30 & 
     1.28 $\pm$ 0.21 &   1 & 63.4 & 42\\
45 &\multicolumn{2}{c}{ 2} & 20 & 
     7.56 $\pm$ 0.94 & 5 & 23.5 & 58 \\
\tableline
\multicolumn{8}{c}
{    Signature: $ee+{\mathrm jets}+$ \met, exclude $80<M_{ee}<105$ }\\ \tableline
20 &\multicolumn{2}{c}{ 2} & 20 & 
     4.84 $\pm$ 0.69 &  5 & 52.5 & 67\\
20 &\multicolumn{2}{c}{ 3} & 20 & 
     1.27 $\pm$ 0.21 &  1 & 63.8 & 40\\
45 &\multicolumn{2}{c}{ 2} & 20 & 
     3.03 $\pm$ 0.48 &   3 & 64.0 & 60\\
{\bf 45}
 &\multicolumn{2}{c}{{\bf 3}} & {\bf 20} & 
     0.93 $\pm$ 0.17 &  0 &  39.5 &  31 \\
45 &\multicolumn{2}{c}{ 3} & 30 & 
     0.80 $\pm$ 0.16 &   0 & 44.9 & 31 \\ \tableline
\multicolumn{8}{c}
{    Signature: $\mu\mu+{\mathrm jets}+$ \met }\\ \tableline
20 &\multicolumn{2}{c}{ 2}  & 20 & 
     1.61 $\pm$ 0.26 &   3 & 22.1 & 68\\
20 &\multicolumn{2}{c}{ 3} & 20 & 
     0.37 $\pm$ 0.10 &  2 & 5.6 & 66\\
{\bf 20}
&\multicolumn{2}{c}{{\bf 2}} & {\bf 30} & 
     0.75 $\pm$ 0.19 &  2 & 17.6 & 60 \\
20 &\multicolumn{2}{c}{ 2} & 40 & 
     0.53 $\pm$ 0.16 &  1 & 40.4 & 46\\
45 &\multicolumn{2}{c}{ 2} & 20 & 
     1.28 $\pm$ 0.24 &  3 & 14.2 & 71\\
45 &\multicolumn{2}{c}{ 3} & 40 & 
     0.12 $\pm$ 0.06 &  1 & 11.4 & 50\\
\tableline
\multicolumn{8}{c}
{    Signature: $e\mu+{\mathrm jets}+$ \met }\\ \tableline
 $\mu$ & $j_1$ & $N_{\mathrm jets}$ & \met & Background & Data & Prob.
 (\%)  & $(\epsilon\sigma)_{\mathrm lim}$  \\ 
\tableline
4 & 20 & 2 & 20 & 
     6.30 $\pm$ 1.04 & 6 & 55.9 & 73\\
4 & 20 & 3 & 20 & 
     1.75 $\pm$ 0.31 &  1 & 47.6 & 41\\
4 & 45 & 2 & 30 & 
     1.97 $\pm$ 0.47 &  2 & 57.2 & 52 \\
4 & 45 & 3 & 30 & 
     0.70 $\pm$ 0.16 &  0 & 49.7 & 31\\
10& 45 & 2 & 20 & 
     1.79 $\pm$ 0.49 &  2 & 52.0 & 53 \\
{\bf 10}
& {\bf 45} & {\bf 3} & {\bf 20} & 
     0.46 $\pm$ 0.14 &  1 & 36.3 & 47 \\
10 & 45 & 2 & 30 & 
     1.35 $\pm$ 0.44 &  0 & 25.9 & 31 \\
10& 45 & 3 & 30 & 
    0.41 $\pm$ 0.13 &  0 & 66.4 & 31 \\
\end{tabular}
\caption{
Representative results for all  signatures. 
For $ee$, $E_T(e_1)>17$~GeV and 
$E_T(e_2)>15$~GeV. For $\mu\mu$, the requirements were 
10 and 20~GeV.
For $e\mu$, each channel required
$E_T(e)>17$~GeV, and $E_T(\mu)$ as specified, $\mu$. 
For all signatures, the leading jet $E_T$ is $j_1$,
and we required $N_{{\mathrm jets}}$ with $E_T>20$~GeV. 
The uncertainty on the 
background  is the sum in quadrature of systematic and statistical contributions.
The probability is for the background to fluctuate to
produce the number of observed events. $(\epsilon\sigma)_{\mathrm lim}$ is the 95\%
C.L. exclusion on the
product of the total cross section, branching ratio, and all 
efficiencies, in fb.
\label{tb:results}}
\end{center}
\end{table}

%% file: dilepton_prl.bbl
\begin{references}
\bibitem{theory} 
P.~Nath {\it et al.}, {\it Applied N = 1 Supergravity},
ICTP Series in Theoretical Physics Vol. 1 (World Scientific,
Singapore, 1984); 
H.~Nilles, Phys. Rep. {\bf 110}, 1 (1984);
H.~Haber and G.~L.~Kane, Phys. Rep. {\bf 117}, 75 (1985);
X.~Tata, in {\it The Standard Model and Beyond},
edited by J. Kim (World Scientific, Singapore, 1991).
\bibitem{lsp}
For the models under consideration, $\tilde{\chi}^0_1$, the lightest
 neutralino,
is the LSP over most of the mSUGRA parameter
space.
\bibitem{thesis} R. J. Genik II, Ph.D. Thesis, Michigan State University, 
{\tt http://www-d0.fnal.gov/results/publications\_talks\-/the\-sis/thesis.html},
 unpublished (1998).
\bibitem{ref:met} B. Abbott {\it et al.} (D\O\ Collaboration), 
 Phys. Rev. Lett. {\bf 82}, 29 (1999). 
\bibitem{sugra}
L.~E.~Iba\~{n}ez, C.~Lopez and C.~Mu\~{n}oz, Nucl. Phys. B {\bf 256},
 218 (1985);
M.~Drees and M.~M.~Nojiri, Nucl. Phys. B {\bf 369}, 54 (1992);
H.~Baer and X.~Tata, Phys. Rev. D {\bf 47}, 2739 (1993);
G.~L.~Kane {\it et al.}, Phys. Rev. D {\bf 49}, 6173 (1994).
\bibitem{ref:d0} S. Abachi {\it et al.} (D\O\ Collaboration), Nucl. Instr.
 and Methods in
Phys. Res., 
A {\bf 338}, 185 (1994). 

\bibitem{ref:jes} B. Abbott {\it et al.} (D\O\ Collaboration), Nucl. Instr.
 and Methods in Phys. Res., A {\bf 424}, 352 (1999).


\bibitem{ref:topprd} B. Abbott  {\it et al.} (D\O\ Collaboration), 
Phys. Rev. D {\bf 58}, 052001 (1998).
\bibitem{ref:topxsect} B. Abbott  {\it et al.} (D\O\ Collaboration), 
Phys. Rev. D {\bf 60}, 012001 (1999).
\bibitem{ref:wxsprd} B. Abbott  {\it et al.} (D\O\ Collaboration), 
Phys. Rev. D {\bf 60}, 052003 (1999).

\bibitem{ref:spythia} 
S. Mrenna, Comput. Phys. Commun. {\bf 101}, 232, (1997). 
{\sc spythia} is a superset of {\sc pythia} 5.7 \cite{ref:pythia},
 allowing us to
    generate SM backgrounds. We incorporate routines
    developed by Kolda\cite{ref:cmssm} to generate the mSUGRA
    model spectrum. This spectrum is passed to {\sc spythia} using the general
    MSSM option.
\bibitem{ref:pythia} T. Sj\"{o}strand, CERN-TH.7112/93, unpublished (1993).
\bibitem{ref:cmssm} G. L. Kane, C. Kolda, L. Roszkowski, and J. Wells,
Phys. Rev.  D {\bf 49}, 6173 (1994). 
 
\bibitem{ref:sig} J. Linnemann, in Proc. of
Comput. in High Energy Phys., Rio De
Janiero, 1995  (World Scientific, Singapore, 1996).

\bibitem{ref:lep1} 
R.~M.~Barnett
{\it et al.}, Particle Data Group, Phys. Rev. D  
{\bf  54}, 165 (1996).

\bibitem{ref:lep2}R. Barate {\em et al.} (ALEPH Collaboration), CERN-EP/99-014, 
submitted to  Eur. Phys. J. (1999). G. Abbiendi {\em et al.}
 (OPAL Collaboration),
CERN-EP-99-123, submitted to  Phys. Lett. B (1999). 

\end{references}
